\documentclass[aps,prl,twocolumn,groupedaddress, showpacs]{revtex4}
\usepackage{graphicx}

\begin{document}

\title{Human Time-Frequency Acuity Beats the Fourier Uncertainty Principle}
\author{Jacob N. Oppenheim}
\author{Marcelo O. Magnasco}
\email[]{magnasco@rockefeller.edu}
\affiliation{Laboratory of Mathematical Physics, Rockefeller University, New York, New York 10065}

\date{\today}

\begin{abstract}
The time-frequency uncertainty principle states that the product of the temporal and frequency extents of a signal cannot be smaller than $1/(4\pi)$. We study human ability to simultaneously judge the frequency and the timing of a sound. Our subjects often exceeded the uncertainty limit, sometimes by more than tenfold, mostly through remarkable timing acuity. Our results establish a lower bound for the nonlinearity and complexity of the algorithms employed by our brains in parsing transient sounds, rule out simple ``linear filter" models of early auditory processing, and highlight timing acuity as a central feature in auditory object processing.\end{abstract}


\maketitle

Fourier transformation turns signals ``inside out", in the sense that low frequencies dictate what happens at long times, while high frequencies create fine temporal detail. This property is demonstrated by Fourier's uncertainty theorem, which states that considering the absolute value squared of a signal $x(t)$ as a probability distribution in time,
\begin{equation} \label{TimeDefEq}
P(t)={|x(t)|^2\over{\int_{-\infty}^\infty|x(t')|^2 dt'}	}			
\end{equation}
and the absolute value squared of its Fourier transform $\tilde{x}(f)$ as a distribution in frequency, 
\begin{equation} \label{FreqDefEq}
P(f)={|\tilde{x}(f)|^2\over \int_{-\infty}^\infty|\tilde{x}(f')|^2 df'}	
\end{equation}			
then the product of the standard deviations 
\begin{equation}
\Delta t=\sqrt{ var(t) } {\ \ \rm and\ \ } \Delta f= \sqrt{ var(f)} 
\end{equation}
is bounded from below [1]: 
\begin{equation}
\Delta t \Delta f \ge {1\over 4	\pi }
\end{equation}
whence it is inferred that short signals require many frequencies for their representation. 

The theorem refers to the original signal and its Fourier transform. In time-frequency analysis one attempts to describe a signal in the two-dimensional time-frequency plane, akin to a musical score where time is the horizontal axis and frequency the vertical axis. Here the uncertainty principle begets the Gabor limit \cite{Gabor,CohenBook}. 
This remapping emphasizes the uncertainties as a property of the transform itself, rather than the the signal. In time-frequency analysis, it has been proven that {\em linear operators cannot exceed the uncertainty bound} \cite{CohenBook}. 
Nonlinearity does not by itself confer any acuity advantage, and in fact most nonlinearities are merely distortions and thus deleterious. However, by the above theorem, any carefully-crafted analysis that can beat this limit must necessarily be nonlinear. For instance, precise frequency information can be obtained about a sine wave by measuring the time between two adjacent zeros of the waveform, a clearly nonlinear operation. The nonlinear distributions  can be classified in families according to their degree of nonlinearity or history-dependence, such as the quadratic (Cohen's class) distributions like Wigner-Ville \cite{Wigner} and Choi-Williams \cite{ChoiWilliams}, and higher-order ones, such as multi-tapered spectral derivatives \cite{Thomson,Mitra}, the Hilbert-Huang distribution \cite{EMD}, and the reassigned spectrograms \cite{KoderaVilledary, AugerFlandrin, Flandrin:DR, Fitz, GardnerMagnasco}. 
To understand how they differ we need to make an important distinction between resolution and precision. Resolution refers to our ability to distinguis two objects, while precision refers to our ability to track the parameters of a single object, given prior knowledge it is only one component. This distinction is well-established in optics, where it is known the wavelength of light limits resolution: two glass beads  cannot be resolved  as different in a microscope if they are closer together than a wavelength.  Precision is not limited, since a single bead can be tracked with nanometer accuracy. All the above distributions achieve higher precision than the Gabor limit when applied to isolated signal components, yet give interfering results when two signals are closer together than an uncertainty envelope. Our experimental test is designed to directly measure precision, not resolution.

A key goal in neuroscience is to establish which algorithms the brain uses to process perceptual information. Psychophysics, by establishing tight bounds on the performance of our senses,may rule out entire families of perceptual algorithms as candidates when they cannot achieve the expected performance \cite{ZwickerFastl, GescheiderBook}. 

We shall show below that human subjects can discriminate better, and occasionally much better, than the uncertainty bounds. This categorically rules out any first order operators, such as the standard sonogram, from consideration, and puts a stringent bound on the performance of any candidate algorithm, demonstrating that the nonlinearities in the cochlea constitute are integral to the precision of auditory processing. 

Our results are relevant both in the scientific and technical areas (e.g. \cite{LeChevalier}), as many high-level models of auditory processing assume an underlying representation of the earliest steps in auditory information homologous to a bank of linear filters \cite{Langner, Hohmann}. Others use one implicitly, by estimating receptive fields from the reverse correlation method (revcor) or by projecting auditory signals onto a basis of more ``natural" filters than the Fourier basis.  In many applications such as speech recognition or audio compression (e.g. MP3 \cite{BosiBook}), the first computational stage consists of generating from the source sound sonogram snippets, which become the input to latter stages.  Our data suggest this is not a faithful description of early steps in auditory transduction and processing, which appear to preserve much more accurate information about the timing and phase of sound components \cite{ Casseday, Shamma,GardnerMagnasco} than about their intensity.  

\begin{figure}
\centerline{\includegraphics[width=8cm]{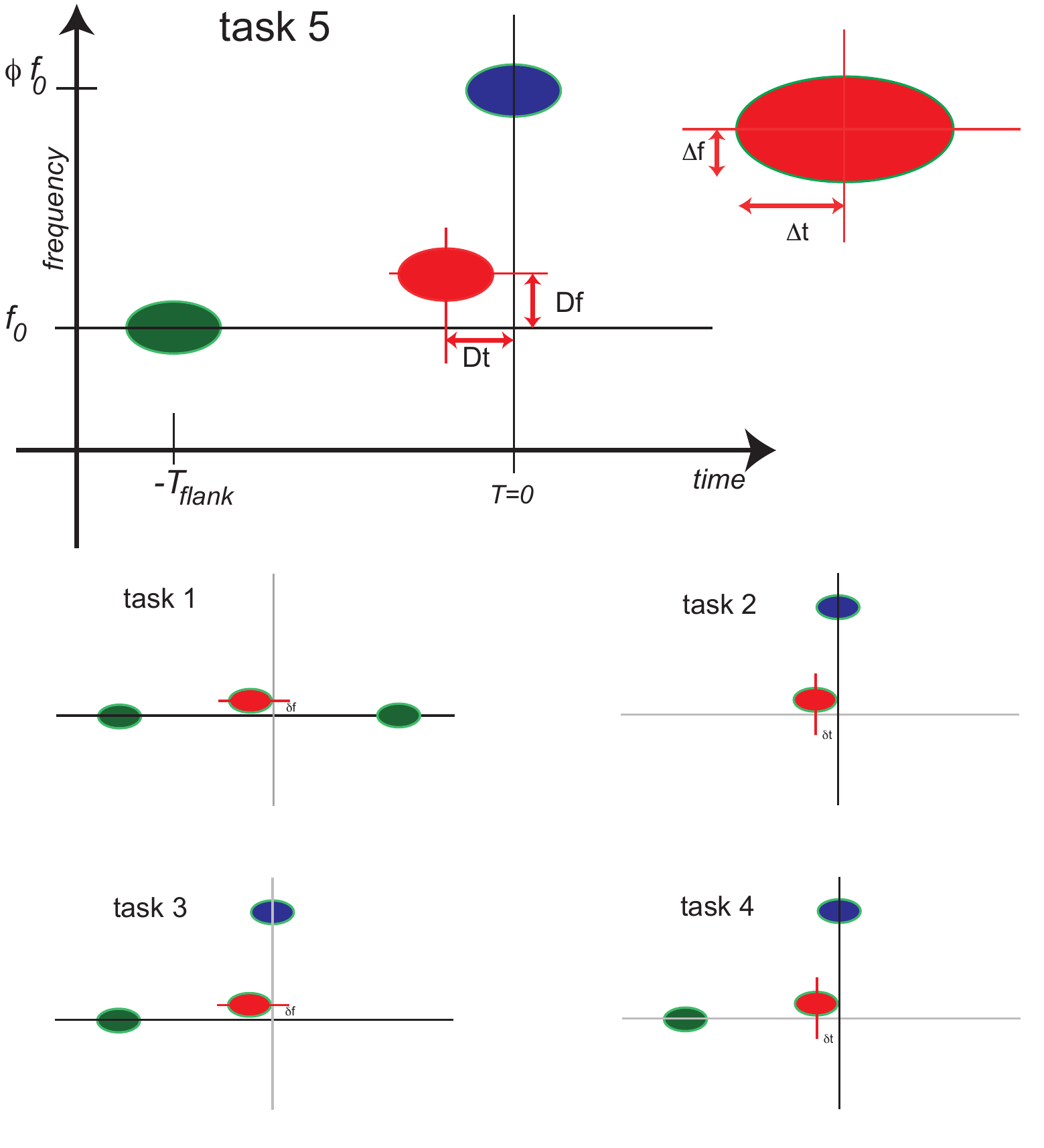}}
\caption{
Stimulus and task. In our final task 5, subjects are asked to discriminate simultaneously whether the test note (red) is higher or lower in frequency than the leading note (green), and whether the test note appears before or after the flanking high note (blue). For each instance of the task, two numbers are generated (Dt and Df) and two Boolean responses (left/right, up/down) are recorded. Tasks 1 through 4 lead to this final task: task 1 is frequency only (uses two flanking notes), task 2 timing only, task 3 is frequency only but with the flanking high note (blue) as a distractor, and task 4 is timing only, with the leading (green) note as a distractor.
\cite{Supplement}.
}\end{figure}

\begin{figure}
\centerline{\includegraphics[width=4.25cm]{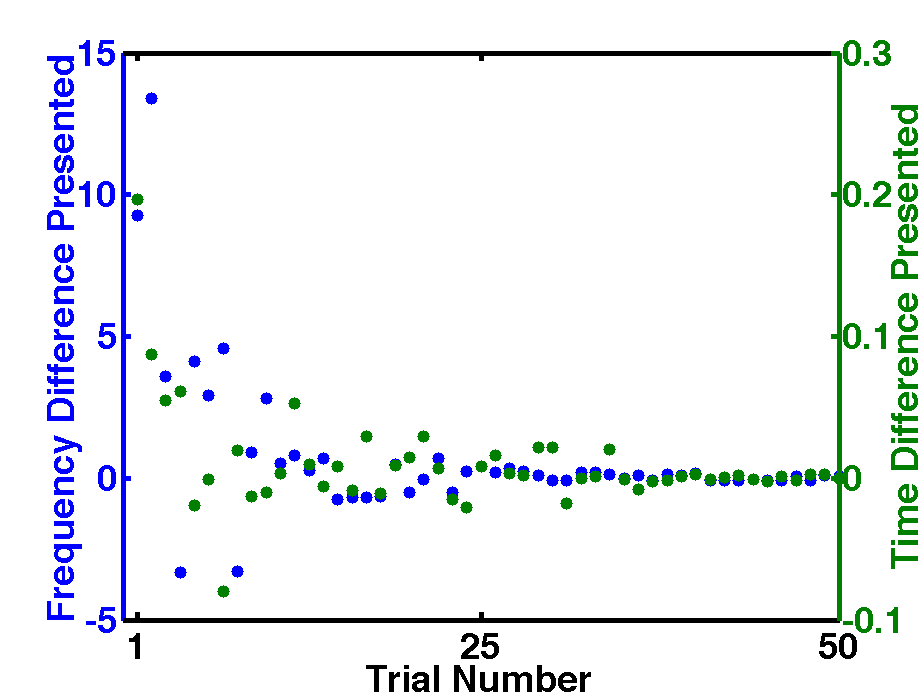}
\includegraphics[width=4.25cm]{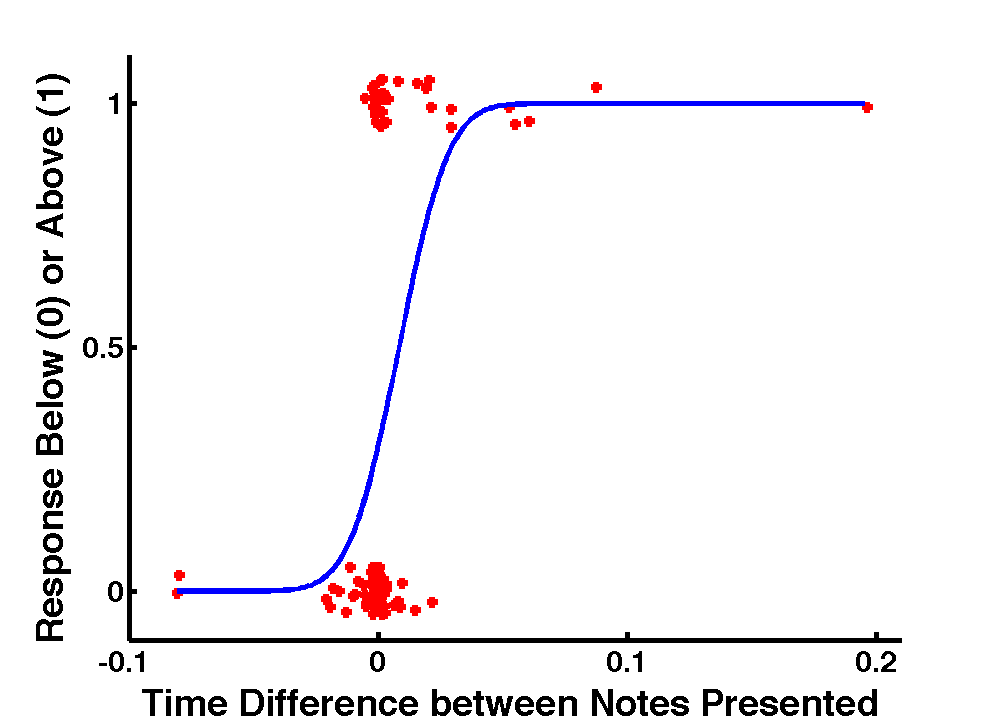}}
\caption{
(A) Measurement strategy. As task 5 proceeds, the numbers Df and Dt are drawn as Gaussian random numbers with variances $pref$ and $multf$. The smaller these variance, the harder the task. The variances are independently controlled by a 2D1U (two down, one up) procedure: when two responses in a row are correct, the variance is reduced and the task is made harder; the variance is increased for every wrong response. This procedure converges to a demanding regime, where the subject makes frequent mistakes, but fewer than 50\%. Data shown from subject {\bf qr3zb} \cite{Supplement}.
(B) Datum definition. We show in red the time responses of subject {\bf qr3zb}; horizontal axis is Dt, vertical axis is 0 (for before) or 1 (for after); we have slightly offset the data by random amounts from 0 or 1 to better visualize the density of points at any given Dt. In blue, the psychometric curve which maximizes the likelihood of the data. The procedure described in 1(b) has converged to a high density of tests around 0, spanning the steepest area of the psychometric curve
}\end{figure}

We shall carefully distinguish between the physical attributes of the stimulus and the analogous psychological quantities. Most relevant will be the distinction between 
$\Delta t$ and $\Delta f$, the {\em physical} uncertainties defined by eqns (\ref{TimeDefEq}, \ref{FreqDefEq}), versus
$\delta t$ and $\delta f$, the {\em psychological} limens of discrimination.
It would be trivial to violate the theorem by using an incorrect definition of $\delta t$ and $\delta f$ or an incorrect evaluation of the bound. The limens are defined to carry the meaning of a standard deviation, so that the actual number is directly comparable to the equivalent physical attribute. It is standard in the literature to define limens through a same-different paradigm. For reasons detailed below, but particularly because same-different is unlike the standard deviation definition of the physical $\Delta t$ and $\Delta f$, we shall operatively define $\delta t$ and $\delta f$ through a two-alternative forced choice above or below paradigm (illustrated by the right panel of Figure 2), and then regress by maximum likelihood the performance data against a psychometric curve in the form of an error function; the standard deviation parameter of this error function is our limen. 

\begin{figure}
\centerline{\includegraphics[width=8cm]{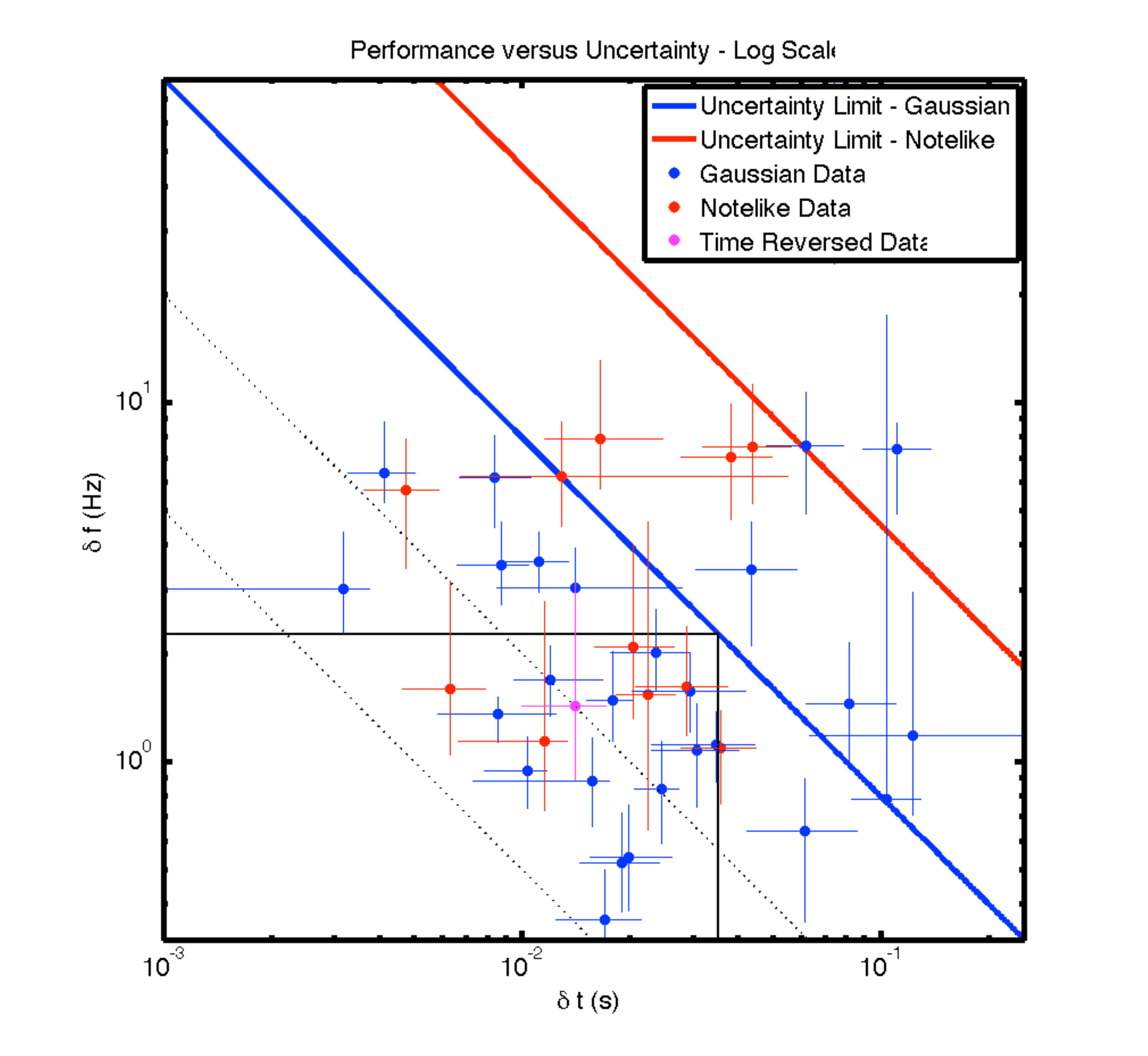}}
\caption{
Figure 2. Summary of main results: discrimination limens for each test. Each round dot is a completion of Task 5 by a subject on an individual day, with at least 100 presentations. There were 12 subjects totaling 26 individual sessions for Gaussian and 12 sessions for notelike tests. Blue denotes Gaussian packet while red denotes notelike. The two solid lines are the locus of the relation $\delta t \delta f=\Delta t \Delta f$; any dots below these curves violate the corresponding uncertainty relation. Error bars in both dimensions were obtained by generating 1000 bootstraps from the raw data and plotting the $25\%-75\%$ quartiles. Raw data provided in Suppl. Table S1.}
\end{figure}

We test for both limens simultaneously, as shown in Figure 1.  Figure 2 demonstrates how we increase difficulty and extract limens of perception. Prior work in the area (e.g.\cite{Ronken, MooreArticle, Freyman}) has always compared measurements of frequency discrimination limens $\delta f$ against the physical temporal duration of the sound packets, $\Delta t$. This is inadequate for our purposes on two grounds: first it treats the quantities in the inequality differently, contradicting the ``spirit" of the uncertainty principle, and second because it fails to verify human ability in an important and ecologically relevant domain, timing acuity.

We use two test stimuli \cite{Supplement}. The first is a Gaussian packet, for which $4\pi\Delta t\Delta f=1$, attaining the bound in the theorem; our study shows that many subjects display limens such that $4\pi\delta t\delta f \ll 1$. In most of the subjects, the overall increase in performance comes from substantial increases in timing accuracy. One of our subjects, {\bf ar4tl}, when tested with notes of $\Delta t=35ms$ attained a limen of $\delta t=3ms$, while frequency performance degraded, $\delta f> \Delta f$. In our second test we use a wave packet with a note-like envelope characterized by a rapid rise and a slow exponential decay. Such envelopes are sub-optimal according to the uncertainty principle, having a product $4\pi\Delta t \Delta f$ greater than unity; in our case, $4\pi\Delta t\Delta f=5.7079$. However the performance of our subjects on such packets is just as good, if not better, than on the Gaussian packet; yielding broad implications for understanding early auditory processing. 

The results from task 5 are summarized in Figure 3 and available in \cite{Supplement}. Each dot corresponds to a simultaneous limen measurement as outlined above. Some subjects performed several different measurements, never on the same week. Two extremes are worth discussing in detail. The lowest blue dot at the bottom center of the plot displays the greatest violation of the principle in our records, by a factor of about 13. The subject {\bf qr3zb} displayed in equal measure a marked increase in frequency acuity as well as temporal acuity, and hence the measurement is below and to the left of the physical values of $\Delta t$, $\Delta f$ for the Gaussian note (indicated by the black lines). The subject is a professional musician. The second point to consider is the leftmost point, at the center left of the diagram, from subject {\bf ar4tl}. This is the smallest $\delta t$ limen in our records; at 3 ms, the subject was able to discriminate the relative timing of two notes a factor of 13 better than their widths; it should be noted that 3ms is barely more than a single period of the test note, 2.27ms. However this subject was unable to estimate frequency better than its physical extent, which is indicated by the dot being above the black line indicating the Gaussians $\Delta f$, so overall this measurement beats the uncertainty principle only by a factor of 10. The subject is an electronic musician who microcomposes and works in precision sound editing. 

We can now examine some implications of these data. First, even though the notelike packet's uncertainty product is substantially above the minimum, subjects seem to be able to discriminate with it just as well as with the Gaussian packet, leading to two measurements (red dot at the bottom of the graph and red dot on the black horizontal line) that beat relative uncertainty by a factor of 50: $\delta t\delta f \approx (1/50) \Delta t\Delta f$, and absolute uncertainty by a factor of 10: $4\pi\delta t\delta f\approx (1/10)$. Therefore we may conclude that a larger uncertainty product of the test note does not necessarily affect the subjects' acuity.  Second, for the Gaussian (blue) data, the plot shows a number of different strategies that subjects use to discriminate, with a remarkable spread: from those who do not achieve the physical limits in either dimension (1), those who have better frequency but worse timing (4), those with better timing and worse frequency (10), and those who have both better timing as well as better frequency discrimination than the physical values (8). While the number of measurements in each category undoubtedly reveals the underlying bias of our subject population, the fact that there are many strategies should be robust. However, there is a noticeable shift of the cloud to the left of the reference notes, so that we can see on median the subjects perform twice as well in timing discrimination as the physical value: $80\%$ of the Gaussian data and $100\%$ of the notelike data lie on the $\delta t<\Delta t$ halfplane. 

It is important to stress where the difficulty of the task lies.  Our preliminary testing included non-musicians, who where often close in performance to musicians on tasks 1 and 2 (separate time and frequency acuity), but then found tasks 3 and 4 hard, while musicians, trained to play in ensembles, found them easy. 

We further found that composers and conductors achieved the best results in task 5, consistently beating the uncertainty principle by factors of 2 or more, whereas performers were more likely to beat it only by a few percentage points.  After debriefing subjects, it appears that the necessity of hearing multi-voiced music (both in frequency and in time) in one's head and coaching others to perform it led to the improved performance of conductors and composers. 

Early last century a number of auditory phenomena, such as residue pitch and missing fundamentals, started to indicate that the traditional view of the hearing process as a form of spectral analysis had to be revised. In 1951, Licklider \cite{Licklider} set the foundation for the temporal theories of pitch perception, in which the detailed pattern of action potentials in the auditory nerve is used \cite{deCheveigne,HCC}, as opposed to spectral or place theories, in which the overall amplitude of the activity pattern is evaluated without detailed access to phase information. The groundbreaking work of Ronken \cite{Ronken} and Moore \cite{MooreArticle} found violations of uncertainty-like products and argued for them to be evidence in favor of temporal models. However this line of work was hampered fourfold, by lack of the formal foundation in time-frequency distributions we have today, by concentrating on frequency discrimination alone, by technical difficulties in the generation of the stimuli, and not the least by lack of understanding of cochlear dynamics, since the active cochlear processes had not yet been discovered. Perhaps because of these reasons this groundbreaking work did not percolate into the community at large, and as a result most sound analysis and processing tools today continue to use models based on spectral theories. We believe it is time to revisit this issue. 

We have conducted the first direct psychoacoustical test of the Fourier uncertainty principle in human hearing, by measuring simultaneous temporal and frequency discrimination. 
Our data indicate that human subjects often beat the bound prescribed by the uncertainty theorem, by factors in excess of 10. This is sometimes accomplished by an increase in frequency acuity, but by and large it is temporal acuity that is increased and largely responsible for these gains. Our data further indicate subject acuity is just as good for a note-like amplitude envelope as for the Gaussian, even though theoretically the uncertainty product is increased for such waveforms. Our study directly rules out many of the simpler models of early auditory processing, often used as input to the higher-order stages in models of higher auditory function. 
Of the plethora of time-frequency distributions and auditory processing models that have been studied, only a few stand a chance of both matching the performance of human subjects  and be plausibly implementable in the neural hardware of the auditory system(e.g.\cite{EMD,HCC,Mitra, GardnerMagnasco}, with the reassignment method having the best comparative temporal acuity. 
Elucidation of which mechanism underlies our subjects auditory hyper acuity is likely to have wide-ranging applications, both in fields where matching human performance is an issue, such as speech recognition, as well as those more removed, such as radar, sonar and radio astronomy.  
\begin{acknowledgments}
 We wish to thank Mayte Suarez-Farinas and Maurizio Pellegrino for their algorithmic and psychophysical expertise, and Tim Gardner for valuable discussions. Supported in part by NSF grant EF-0928723. 
\end{acknowledgments}

\end{document}